\documentclass[11pt]{article}

\newcommand{\lbl}[1]{\qquad\qquad\framebox{\scriptsize #1} \label{#1}}
\newcommand{\bib}[1]{\bibitem{#1} \qquad\framebox{\scriptsize #1}}

\usepackage{amsxtra}
\usepackage{amsmath}
\usepackage{graphics}

\def\be{\begin{equation}}
\def\bea{\begin{eqnarray}}
\def\ee{\end{equation}}
\def\eea{\end{eqnarray}}
\def\bc{\begin{center}}
\def\ec{\end{center}}
\def\ba{\begin{array}}
\def\ea{\end{array}}
\def\bt{\begin{tabular}}
\def\et{\end{tabular}}
\def\bq{\begin{quote}}
\def\eq{\end{quote}}
\def\bi{\begin{itemize}}
\def\ei{\end{itemize}}

\def\1{{\pi}}
\def\v1{{\varpi}}
\def\q{{\gamma}}

\def\e{{\epsilon}}

\def\O{\mbox}

\def\a{{\alpha}}

\def\f{\frac}

\def\L{{\Omega}}

\def\x{{\chi}}
\def\X{~{\equiv}~}

\def\v{\\[.1in]}
\def\B{{\infty}}

\def\M{{\partial}}

\def\rb{\right]}
\def\lb{\left[}

\def\'{{\prime}}
\def\bthm{\begin{thm}}
\def\ethm{\end{thm}}
\def\blm{\begin{lm}}
\def\elm{\end{lm}}
\def\bprop{\begin{prop}}
\def\eprop{\end{prop}}

\def\bs{\begin{slide}}
\def\es{\end{slide}}
\def\bbe{\begin{boldequation*}}
\def\ebe{\end{boldequation*}}

\def\V{\hat{V}}
\def\t{\hat{t}}
\def\T{\hat{T}}
\def\b{\hat{B}}
\def\L{$\hat{L}$ }
\def\m{$\hat{M}$ }
\def\b{\hat{B}}
\def\H{\hat{H}}
\def\E{\hat{E}}
\def\D{\Delta}
\def\X{\equiv}

\begin{document}

\newtheorem{thm}{Theorem}
\newtheorem{lm}{Lemma}
\newtheorem{prop}{Proposition}

\title{Special and general relativity extended to allow for a variable light speed}
\author{Robert C. Fletcher\footnote{Mailing address: 1000 Oak Hills Way, Salt Lake City, UT 84108.  robert.c.fletcher@utah.edu}
\\
{\small Bell Laboratories (ret)}\\
{\small Murray Hill, New Jersey.}}  
\date{}

\maketitle

\bq {\small \bc {\bf Abstract}\ec

There have been a number of papers proposing that the light speed of a homogeneous and isotropic universe is variable.  This paper outlines a simple way that the vectors and tensors of special relativity can be extended to allow such a variable light speed and show how the field equation of general relativity can also be extended with particular application to such a universe. 
}\eq

\tableofcontents

\section{ Introduction} \lbl{intro}

There have been a number of attempts in the literature to investigate a variable speed of light.  Those mostly 
tried to find a new cosmology to provide alternatives to inflation in the ``standard'' Friedman-Lemaitre-Robertson-Walker (FLRW) cosmology\cite{R}\cite{Wa} to resolve horizon and flatness problems\cite{Bas}\cite{HSK}\cite{Mag1}\cite{Mag3}.  I have not tried to impose a variable rate in order to achieve any goal, but have attempted to extend relativity
with a minimum of changes in order to accommodate the possibility of any variable light speed in a homogeneous and isotropic universe. G F R Ellis and J-P Uzan\cite{EU} have made an elegant description of what is required of a variable light speed.  I believe the present paper is consistent with their requirements if we assume, in keeping with an aim for simplicity, that the three different ``light'' speeds they identify are all the same $c(t)$.

The aim of the present paper is to outline a way that not only Lorentz, but all of special (SR) and general relativity (GR) can be extended to allow such a variable light speed.
 The extended Lorentz transform for local coordinates is derived from the basic assumption of relativity that the light speed $c$ is the same for all moving observers at the same space-time point even though the light speed and their relative velocity $V$ may vary.  To form SR vectors and tensors we use a differential construct 
$d\T = cdT$ from physical time $T$\cite{Mag3} and a dimensionless velocity $\hat{V} = V/c$.  
In addition we propose that the rest mass of a particle varies so as to keep its rest energy constant.  This seems reasonable inorder to eliminate the need for an external source or sink of energy for the rest mass.
These  assumptions simplify the construction of SR vectors and conserves the stress-energy tensor of an ideal fluid.  For GR, we propose the standard GR action S, but use the extended stress-energy tensor and allow the gravitational constant $G$ to vary with $c$.  The variable light speed is introduced in the line element that determines the space-time curvature.
 
We will use the notation $t$ for time when the light speed is $c(t)$, as it must be for a uniform and isotropic universe if it is to be variable.  Then $\t$ can be a transform from $t$: $\t = \int{c(t)dt}$. 
The GR curvature tensor is derived from a line element that typically has the time $t$ appearing in the combination of $c(t)dt$ that would require the tensor to contain the derivatives of $c$.  
The use of $\t$ instead of $t$ eliminates these derivatives without changing the relations of the components of the tensors, and also allows all the relations of curvelinear coordinates used for constant $c = 1$ to be retained.    
Then, from a solution with $\t$ the observable physical $t$ can be found with a transform from $\t$ to $t$.

\section{The extended Lorentz transform and Minkowski metric}\lbl{LT}

Let us consider two physical frames moving with respect to each other.  The first frame $(S)$ will have clocks and rulers whose readings we will represent by $T$ and $x$.  The second frame $(S^*)$ will move in the $x$ direction at a velocity of $V = (\M x/\M T)_{x^*}$ as measured by $T$ and $x$ and will  have clocks and rulers whose coordinates we will represent by $T^*$ and $x^*$.  The velocity of the first frame will be $V^*$ as measured by $T^*$ and $x^*$.  We assume that the light speed, even though variable, is the same as measured on both frames at the same space-time point.  We also allow $V$ to be variable.

 In order for $S$ to measure the small separation of points $\D x^*$ on $S^*$, $S^*$ sends two simultaneous ($ \D T^* = 0$) signals as measured on its clocks, one at the beginning of $\D x^*$ and the other at the end.  $S$ measures the space between the signals as $\D x$, but does not see these signals as simultaneous.  The far end signal is delayed by $\D T$ over the near end signal for this reason.  $S$ measures $\D x^*$ to be the distance $\D x$ reduced by the distance that $S^*$ has traveled in the time $\D T$ after $S$'s simultaneity ($\D T = 0 $) with the near end, i.e., $\D x - V \D T$.  Since we are looking for linear relationships, we assume that the $S^*$ measure of $\D x^*$ is proportional to the $S$ measure:
\be 
\D x^* = \alpha (\D x - V \D T)
\lbl{lt2} \ee
where we have allowed $\alpha$ and $V$ to be varying, but approach a constant value for small $\D$ 's.
  We also assume that for the two Cartesian directions $\D y$ and $\D z$ perpendicular to the motion along $x$ that the $S^*$ and $S$ coordinates are the same 
\be \D y = \D y^*, \D z = \D z^*
\lbl{lt1}\ee  
and that the time $T$ does not depend on $y$ or $z$.  $\alpha$ will be determined from the assumption that the light speed is the same on all moving frames. We will adapt the analysis of Bergmann\cite[pp33-36]{brg} to a variable light speed.  Choosing the point of origin so that $\D T$ and $\D T^*$ vanish when $\D x$ and $\D x^*$ vanish, we expect that $\D T^*$ will be a linear function of $\D T$ and $\D x$: 
 
\be 
\D T^* = \q \D T + \zeta \D x
\ee 
where $\alpha, \q$, and $\zeta$ are slowly varying functions that approach a constant for small $\D$'s.  We will now determine their values.

We assume that the light speed can be variable, but in small intervals of time and distance it will be almost constant.  It will
have the same values in $S^*$ as in $S$ at the same space-time point.  For light moving in an arbitrary direction, each measures the light speed $c$ as the change in distance divided by the change in time of its own coordinates:
\begin{align}
\D x ^2 + \D y ^2 + \D z^2 & = c^2  \D T^2, \lbl{lt3'}\\
\D x{^*} ^2 + \D y{^*} ^2 + \D z{^*}^2 & = c^2  \D T{^*}^2, \lbl{lt3}
\end{align}
where we have chosen an origin where all the $\D$'s vanish.  
 By using Eqs \ref{lt1} and \ref{lt2} in Eq \ref{lt3}, we can eliminate the starred items to get
\be 
 \alpha ^2 (\D x - V \D T)^2 + \D y^2 + \D z^2 = c^2(\q \D T + \zeta\D x)^2. 
\ee
We can rearrange the terms to obtain
\be 
 (\alpha ^2 - c^2\zeta ^2)\D x^2 - 2(V\alpha ^2+c^2 \q \zeta) \D x \D T + \D y^2 +\D z^2 = (c^2\q ^2 -V^2 \alpha ^2)\D T^2.
\ee 
If we compare this to eq \ref{lt3'} we get
\begin{align}
c^2 \q ^2 -V^2 \alpha ^2 & = c^2 \\
\alpha ^2 - c^2 \zeta ^2 & =1 \\
V \alpha ^2 + c^2 \q \zeta & = 0
\end{align}
We can solve these three equations for the three unknowns $\alpha, \q$, and $\zeta$:
\begin{align}
\q^2 & = \f {1}{1-V^2/c^2} \\
\zeta & = \f {1-\q^2}{\q V} =- \f {\q V}{c^2} \\
\alpha ^2 & = -\f{c^2 \q \zeta}{V} = \q^2
\end{align}

Thus in the differential limit of $\D 's$ going to zero, we write them as differentials, so the relation of differentials becomes 
\begin{align}
dT^* & = \q(dT-Vdx/c^2), \lbl{lt} \\
dx^* & = \q(dx - VdT),
\end{align} 
 By inverting this we get
\begin{align}
dT & = \q(dT^*+Vdx^*/c^2), \lbl{SR} \\
dx & = \q(dx^* + VdT^*). \lbl{sr'}
\end{align}
so $V{^*} = -V$ as you would expect.

This is the same as for a constant $c$, except here $c$ has been allowed to vary.  

We define a line element $ds$ by the relation
\be 
ds^2 \equiv c^2 dT^2 - dx^2 - dy^2 - dz^2
\lbl{m1}\ee
If we substitute eqs \ref{lt1}, \ref{SR} and \ref{sr'} into eq \ref{m1}, the form is the same:
\be 
ds^2 = c^2 dT{^*}^2 - dx{^*}^2 - dy{^*}^2 - dz{^*}^2
\lbl{m2} \ee
That is, the extended world line is invariant in form to changes in coordinates on frames moving at different velocities. The line element is symmetric in the spatial coordinates, so it is valid for motion in any direction.  In polar coordinates this becomes
\begin{align} 
ds^2 & = c^2dT^2 - dR^2 - R^2 d\theta^2 -R^2 sin^2\theta d \phi ^2 \lbl{sr3'} 
  \end{align}
This is the Minkowski line element (\m) extended to allow for a variable light speed.

Notice that if we divide eqs \ref{m1} and \ref{m2} by $c^2$  the two equations still have identical forms, so that the differential time $ d \tau \equiv ds/c$ is also invariant in form to \L transforms.  Since $d\tau = dT$ for constant spatial coordinates, $\tau $ is the time on a clock moving with the frame.  

This derivation has depended on a physical visualization so that we assume that differentials that represent physical time and radial distance must have \m metric for their time and distance differentials and an extended Lorentz transform \L to other colocated physical differentials of time and distance on a frame moving at a velocity $V$.  We will call such differentials physical coordinates.  Time and distance coordinates that do not have these relations will not be physical; one or the other may be physical, but not both unless they have a \m metric.

The extended Lorentz transform \L  can be written in a symmetric form using $d\T \equiv cdT$ and $\V \equiv V/c$ with the velocity in the $R$ direction (as it will be in a homogeneous and isotropic (FLRW) universe):
\begin{align}
d\T^* & = \q(+d\T - \V dR), \lbl{lt5}\\
dR^* & = \q(-\V d\T +dR).
\end{align}

In general for a varying $c$, $\T$ is not a transform from $T$ alone, although, as we have shown in eq \ref{lt5}, we can use the construct $d\T = cdT$ to describe the \L transform.
In a FLRW universe for events in the radial direction measured by the variables $(t,\x)$, if $c$ is variable, it is a simple function of $t$ since homogeneity in space makes it independent of $\x$. In this case $\t$ can be a transform from $t$ alone.  

\section{Extended SR particle kinematics}\lbl{sr}

In this section I will outline the way vectors and tensors can be defined when the light speed is variable.   
In Cartesian coordinates, let $dx^1,dx^2,dx^3 = dx,dy,dz $, and $dx^4 = d\T = cdT$.  The \m metric then becomes
\be
	ds^2 = \eta_{\mu \nu}dx^\mu dx^\nu,
\lbl{sr1"} \ee
where $\eta_{\mu \nu} = (-1,-1,-1,+1)$ for $\mu = \nu$, and zero for $\mu \ne \nu$. The velocity $\dot{x}^\mu$ is $dx^\mu/d\T = V^\mu /c$ with $\dot{x}^4 = 1$.  (The dot represents the derivative with respect to $d\hat{T}$).  The world velocity becomes 
\be 
{U}^\mu = dx^\mu/ds = \q \dot{x}^\mu.
\ee
  $\dot{x}^\mu$ and $U^\mu$ are therefor dimensionless.  In order to make the rest mass energy constant, we define $\hat{m} = mc^2$ and the extended energy-momentum vector as 
\be
	P^\mu = \hat{m}U^\mu = \hat{m} \q \dot{x}^\mu, 
\lbl{sr2}\ee
so that $P^4 = \hat{m} \q = E$.  If $p$ is the magnitude of the physical momentum ($\q mV$), the EP vector magnitude is $E^2 - c^2 p^2 = \hat{m}^2 $.  It has units of energy rather than momentum or mass.

The \L transform for the components of the EP vector is
\be
	E^* = \q(E- \V pc).
\lbl{sr3} \ee
For photons, $\hat{m} = 0$, so $E = h \nu$ and $p = h/\lambda = h\nu /c$, and the Lorentz transform is 
\be
	\nu^* = \q \nu (1-\V).
\lbl{sr4} \ee
This is the familiar relativistic Doppler effect.  

The force vector becomes 
\be
	F^{\mu} = \f {dP^{\mu}}{ds} = \hat{m} A^{\mu} = \hat{m} \f {dU^{\mu}}{ds} .
\lbl{sr5'}\ee
The first three components $F^i/\q$ will
 be the force $f^i$ felt by an object of mass $m$ when the light speed is $c$; ($i$ represent the three spatial coordinates).  In taking the derivative of $P^i$, we are implying that
$mc\f{d(\q V/c)}{dT}$ is more fundamental in determining the physical force than $m \f {d(\q V)}{dT}$ when the light speed is variable.
We can express the gravitation force in the usual way as $mg^i$, where $g^{i} = A^{i}c^2/\q$.  $cF^4/\q$ is the rate of work $f^i V^i$ required to change the rate of change of energy $d(\q \hat{m})/dT$.
All these world vectors are invariant to the \L transform and the \m line element.  They become the usual vectors when $c$ is constant.

\section{Extended analytical mechanics}

We will next show how the Euler-Langrange equations apply to extended particle kinematics\cite{brg}.  For a mechanical system with conservative forces in (n+1)-dimensional space whose differentials are
$(dx^i,d\T)$, the action $S$ is
\be 
S = \int{\hat{L}}ds.
\ee 
Minimizing $S$ gives relations for $\hat{L}$, the Lagrangian.   With no force acting, we will use

 
 \be 
\hat{L} = \hat{m}\sqrt{\eta _{\mu \nu}U^\mu U^\nu},
\ee
so the momenta are
\be 
P_\mu = \f {\M \hat{L}}{\M U^\mu} = \f {\hat{m} \eta_{\mu \nu} U^\nu }{\sqrt{\eta _{\mu \nu}U^\mu U^\nu}}.
\ee 
The root in this equation has the value 1 which makes it possible to solve it for $U^\mu $
\be 
U^\mu =\f { \eta ^{\mu \nu} P_\nu }{\hat{m}\sqrt{\eta _{\mu \nu} U^\nu U^\nu}} = \f {P^\mu}{\hat{m}}, 
\ee 
consistent with eq \ref{sr2}.

So,
\be 
U^\mu P_\mu = \f{\eta^{\mu \nu}P_\mu P_\nu}{\hat{m}}.
\ee 
The Hamiltonian  $\hat{H}$ becomes
\be 
\hat{H} = -\hat{L} + U^\mu P_\mu = -\sqrt{\eta^{\mu \nu}P_\mu P_\nu } + \f{\eta^{\mu \nu}P_\mu P_\nu}{\hat{m}}. 
\ee 
Let $p \equiv \sqrt{\eta^{\mu \nu}P_\mu P_\nu }  = \hat{m}$, so 
\be 
\hat{H} =  \f {p^2}{\hat{m}} - p
\ee 
Thus $\hat{H}$ vanishes,  but its derivative with respect to $P_\mu$ does not:
\begin{align}
U^\mu & = \f {\M \hat{H}}{\M P_\mu} = 2\f{\eta^{\mu \nu}P_\mu } {\hat{m}}
- \f{\eta^{\mu \nu}P_\mu } {p}=  \f {P^\mu}{\hat{m}} \\
\f {dP_u}{ds} & = -\f {\M \hat{H}}{\M x^\mu} = 0,
\end{align}
$P_\mu$ is conserved since we have considered no force acting.

\section{Extended stress-energy tensor for ideal fluid}\lbl{gr}

An ideal fluid can be treated in a similar way.  It is a collection of $n$ particles per unit volume of mass $m$.  We can form a rest energy density function $\hat\rho = n\hat{m}$.  In this case, $\hat{\rho}$ is not constant because $n$ is a a function of time and distance.  
We will use $t$ instead of $T$ to indicate that we are initialy limiting this analysis to $c$ being a function of $t$.  This can be transformed to other frames by a \L transform.  It turns out that $\hat{\rho}$ using $d\t$ and $u^\mu = V^\mu /c(t)$ has much the same properties as $\rho  = nm$ using $dt$ and $V^\mu$ with constant $c$.  

The conservation law for particles in nonrelativistic terms for $n$ flowing at a velocity $V^i = cu^i$ is
\be 
\f { \M n }{\M \t} +  n u^i_{,i} = 0.
\ee
where we have assumed that the differential of $c$ with distance is zero. 
For the conservation of energy we must include the stress forces $t^{ij}dA_j$ operating on the area of the differential volume, like the pressure $p$ where $t^{ij} = p\delta ^{ij}$.  We can convert the area stress forces by Gauss' theorem to a volume change in momentum to give a total 3D energy flux of $cP^i$, where
\be 
P^i = \hat{\rho}u^i + u^j t^{ji}.
\ee
The conservation of the fluid rest energy ($u^i = 0$) then becomes
\be 
\f{ \M \hat {\rho }}{\M t} + \O {div} (cP^i) = 0,
\ee
 or
\be 
\f{ \M \hat {\rho }}{\M \t} + P^i_{,i} = 0.
\lbl{if2} \ee

The Newtonian law linking the rate of change of the generalized velocity $u^i = \f {dx^i}{d\T}$ to the force per unit volume $f^i$ in nonrelativistic terms can be written as
\be 
\hat{\rho} \f {d u^i}{d\t} = f^i 
\ee
We can follow through the steps in any of the standard texts \cite{brg} to obtain the generalized
 stress-energy tensor of an ideal fluid in its rest frame to be
 \[
	 T^{\mu \nu} = T_{\mu \nu} =
	\begin{pmatrix}
		& p
		& 0
		& 0
		& 0 \\[10pt]
		& 0
		& p
		& 0
		& 0 \\[10pt]	
		& 0
		& 0
		& p
		& 0 \\[10pt]
		& 0
		& 0
		& 0
		& \hat{\rho}
 \end{pmatrix}.\] \lbl{if4} 
This is used in Sect \ref{FE}.

This can be generalized for a frame moving at a world velocity $U^\mu$:
\be
	 T^{\mu \nu} = (\hat{\rho} + p)  U^\mu U^\nu  - p\eta^{\mu \nu}.
\lbl{sr5}\ee
One can see that this is the same tensor since in the rest frame of the fluid $\T = \t$, $U^i = 0$, $U^4 = 1$.

The divergence of the stress-energy tensor is the force per unit volume:
\be 
 T^{\mu \nu}_ {\quad,\nu} = F^\mu
\ee 

The conservation of rest energy density (eq\ref{if2}) can then be written:
\be 
 F^\mu =  T^{\mu \nu}_{\quad,\nu}  = 0.
\ee

\section{Extended electromagnetic vectors and tensors}\lbl{em}

We will assume that the light speed that appears in electromagnetic theory (E/M) is the same as appears in relativity theory.  If it were not so, it would be a remarkable coincidence if they were the same today, but different at other times.  The E/M light speed obeys the relation
\be 
c^2 = 1/{\epsilon_0 \mu_0},
\ee 
where $\epsilon_0$ and $\mu_0$ are the electric and magnetic ``constants'' of free space, resp.  If $c$ is variable, then either $\epsilon_0$ or $\mu_0$ or both must vary.

Current measurements with atomic clocks (\cite{Peik}, \cite{SGK}) have achieved an accuracy that indicate the frequency of atomic spectra do not change with time.  Of course, when measured on a frame moving at a different velocity or in a gravitational field, frequency does change.  There are also astronomical indications of a variation in $\a_F$ \cite{Webb}, but these are much smaller than would occur if $c(t)$ changed as calculated in Ref \cite{RCF}).  On an inertial frame, this means that the fine structure constant  $\a_f$ and the Rydberg constant $R_\B c$ (expressed as a frequency) do not change with $c(t)$.

The fine structure constant $\a_f$ in SI units \cite{C} is
\be
	\a _f = \f{e^2}{4\pi \epsilon_0 \hbar c},
\lbl{d9}
\ee
and the Rydberg frequency is
\be
	R_\B c = \a _f ^2 \f{m_e c^2}{4 \pi \hbar} = \f{e^4 m_e}{\epsilon_0 ^2(4\pi \hbar)^3}.
\lbl{d10}
\ee
Because $\a_F$ is dimensionless, the $4\pi \e_0$ is often omitted in the fine structure constant since it is unity in Gaussian coordinates, but it is essential here if we are to consider a variable $c(t)$ for the universe.  

For these to remain constant while keeping $e$, $\hbar$ and $mc^2$ constant requires that
\be
	\epsilon_0(t) c(t) = \f{1}{\mu_0(t) c(t)} \equiv k = \epsilon (t_0) c (t_0), \; \O{a constant}.
\lbl{d11}\ee
$ 1/k = \sqrt{\mu  _0/\epsilon _0} $, the impedance of free space.
This assumption means that the electrostatic repulsion $f$ between two electrons will vary:
\be 
f = -\f {e^2}{\epsilon_0 R^2}.
\lbl{d9'}\ee 

Maxwell's equations in 3D vectors in the rest frame of FLRW with a constant speed of light \cite{Str} are 
\be \ba{rcl}
	\O {curl} E + \f{\M B}{\M t} = 0, \\[.15cm]	
	\O {curl} H - \f{\M D}{\M t} = J, \\[.15cm]
	\O {div} D = \sigma, \\[,15cm]
	\O {div} B = 0, \\[,15cm]
	\O {div} J + \f{\M \sigma}{\M t} = 0.
\ea \lbl{d12}
\ee
Scaler $\phi$ and vector $A$ potentials can be introduced such that
\be \ba{rcl}
	B = \O {curl} A, \\[.15cm]
	E = -\O{grad} \phi - \f{\M A}{\M t},
\ea \lbl{d13} \ee
and the equation for the force on a particle with charge $q$, mass $m$, and velocity $V$ is (see \cite[p118]{brg})
\be
	m\f{d (\q V)}{dt} = q(E + V \otimes B)
\lbl{d14} \ee

With the use of $\t$ and eq \ref{d11} and with the relations $D = \epsilon _0 E, B = \mu_0 H$ for free space, these can be converted 
to exactly the same equations by replacing $t$ by $\t$ and by replacing the field variables by hat variables so that the partial time derivatives of hat variables do not include $\epsilon_0, \mu_0$, or $c$ except in combinations equalling $k$, a constant.  This is accomplished by the following: $\b = k B = \H = H/c$, $\hat{D} = D = \E = \epsilon_0 E$, $\hat{\sigma} = \sigma$, $\hat{J} = J/c$, $\hat{A} = k A$, $\hat{\phi} = c \phi/k$, and $\hat{q} = q$.  Thus, with hat variables and $\t$, Maxwell's equations have only two fields $\hat{E}, \hat{H}$ with no varying coefficients.  
\be \ba{rcl}
	\O {curl} \hat{E} + \f{\M \hat{H}}{\M \t} = 0, \\[.15cm]	
	\O {curl} \hat{H} - \f{\M \hat{E}}{\M \t} = \hat{J}, \\[.15cm]
	\O {div} \hat{E} = \hat{\sigma}, \\[,15cm]
	\O {div} \hat{H} = 0, \\[,15cm]
	\O {div} \hat{J} + \f{\M \hat{\sigma}}{\M \t} = 0.
\ea \lbl{d12}
\ee
The potential equations become
\be \ba{rcl}
	\hat{H} = \O {curl} \hat{A}, \\[.15cm]
	\hat{E} = -\O{grad} \hat{\phi} - \f{\M \hat{A}}{\M \t},
\ea \lbl{d13} \ee
Since they have no coefficients that vary with time, they are \L covariant to frames with $d\hat{T}$ replacing $d\t$ just like the original Maxwell's equations.  Thus, they are valid in every moving frame whose physical time is $T$.

With $\hat{V} = V/c$, and $\hat{m} = mc^2$, The pondermotive equation \ref{d14} becomes
\be
	\hat{m} \f{d(\q {\V})}{d\T} = \f{q}{\epsilon _0}(\hat{E} + \V \otimes \hat{H}).
\lbl{6} \ee
These all become the usual expressions when the speed of light is constant $c = 1$.  

E/M world vectors and tensors can be constructed in the usual way\cite{brg}.  
Thus, the extended covariant potential vector is $\hat{\phi} _\mu = (\hat{A}_i ,-\hat{\phi})$, the extended charge  vector $\hat{\Gamma} ^\mu = (\hat{J}^i ,-\hat{\sigma})$, and the extended covariant E/M field tensor is
\[
	\hat{F}_{\mu \nu} =
	\begin{pmatrix}
		& 0
		& -\H_3
		& +\H_2
		& -\E_1 \\[10pt]
		& +\H_3
		& 0
		& -\H_1
		& -\E_2 \\[10pt]	
		& -\H_2
		& +\H_1
		& 0
		& -\E_3 \\[10pt]
		& +\E_1
		& +\E_2
		& +\E_3
		& 0
\end{pmatrix}.\]

%
The field tensor can be obtained from the curl of the potential vector 
\be
	\hat{F}_{\mu \nu} = \hat{\phi}_{\mu,\nu} -\hat{ \phi}_{\nu,\mu} \ ,
\lbl{sr7} \ee
and Maxwell's equations become the divergence of the field tensor equaling the charge vector \cite[p113]{brg}:
\be
	\hat{F}^{\mu \nu}_{\quad,\nu} = -\hat{\Gamma}^\mu.
\lbl{sr8} \ee
The pondermotive equation for a particle of charge $q$ and mass $m$ becomes a force vector equaling $\hat{m}$ times an acceleration vector:
\be
	\f {q}{\epsilon_0} \hat{F}_{\mu \nu} \hat{U}^{\nu} = -\hat{m} \eta_{\mu \nu} \f{d\hat{U}^\nu}{ds}. 
\lbl{sr9} \ee
The SR stress-energy tensor is
\be 
T^{\mu\nu} = \f {1}{\epsilon_0}[F^\mu_\lambda F^{\nu \lambda} -\f {1}{4}\eta^{\mu \nu}F^{\mu \sigma} F_{\sigma \nu}]
\ee 
For GR with curvilinear coordinates, the stress-energy tensor is
\be 
T^{\mu\nu} = \f {1}{\epsilon_0}[F^\mu_\lambda F^{\nu \lambda} -\f {1}{4}g^{\mu \nu}F^{\mu \sigma} F_{\sigma \nu}]
\ee 
The dimensions of $\hat{E}$,$\hat{H}$, and $F^{\mu \nu}$ are electric charge per unit area, whereas for $T^{\mu \nu}$ it is energy per unit volume.  Because of $\epsilon_0$, both the force and the energy density are dependent on $c(t)$ just like the force between two electrons (eq \ref{d9'}).

\section{The extended metric for a homogeneous and isotropic universe} \lbl{FLRW}

We assume that the concentrated lumps of matter, like stars and galaxies, can be averaged to the extent that the universe matter can be considered continuous, and that the surroundings of every point in space can be assumed isotropic and the same for every point.

By embedding a maximally symmetric (i.e., isotropic and homogeneous) three dimensional sphere, with space dimensions $r$, $\theta$, and $\phi$, in a four dimension space which includes time $t$, one can obtain a differential line element $ds$ \cite[page 403]{We1} such that
\be
	ds^2 = g(t)dt^2 - f(t) \lb \f {dr^2}{1-kr^2} +r^2 d\theta^2 +r^2 sin^2\theta d \phi ^2 \rb,
\ee
where 
\be
	r = \left \{ \ba{l} \sin{\x},~~~k = 1,\\ \x,~~~~k = 0,\\ \sinh{\x},~~~k = -1,
	\ea \right.
\lbl{2"} 
\ee
$k$ is a spatial curvature determinant to indicate a closed, flat, or open universe, resp., and 
\be
d\x^2 \X dr^2/(1-kr^2).
\ee

We let $a(t) \X \sqrt{f(t)}$ be the cosmic scale factor multiplying the three dimensional spatial sphere, so that the differential radial distance is $a(t)d\x$.

The $g(t)$ has normally been taken as $g(t) = c^2 = \mbox{constant}$, so that $c$ is the constant physical light speed and $t$ is the physical time on each co-moving point of the embedded sphere.  In both cases by physical, we mean that their value can represent measurements by physical means like standard clocks and rulers, or their technological equivalents.  In order to accommodate the possibility of $c(t)$ being a function of time, we make $g(t) = c(t)^2$.  The resulting equation for the differential line element becomes a generalized FLRW metric: 
\be
	ds^2 = c(t)^2dt^2 - a(t)^2[d\x^2 + r^2d\omega^2], 
\lbl{1}
\ee 
where $d\omega^2 \equiv d\theta^2 + sin^{2} \theta d\phi^2 $.   
For radial world lines this metric becomes  Minkowski in form
with a differential of physical radius of $a(t)d\x$.

It will be convenient to introduce the time related quantity $\t$, which we will call a generalized cosmic time, defined by
\begin{align}
\t  & \equiv \int_0^{t}{ c(t)dt},\\	t & = \int_0^{\t}{\f{ d\t}{\hat{c}(\t)}},
\lbl{3'}
\end{align}
where $\hat{c}(\t) = c(t)$, and where the lower limit is arbitrarily chosen as $0$.

The line element then becomes
\be
	ds^2 = d\t^2 - a^2(d\x^2 + r^2d\omega^2).
\lbl{3"}
\ee

It should be emphasized that $\t$ itself is a legitimate more general coordinate that makes the metric of FLRW satisfy the Cosmological Principle; the physical time $t$ is a transform from it. $\t$ and its transform to $t$ allows for the physics to apply to a variable light speed.  

\section{Extended GR field equation} \lbl{FE}

We assume the standard action of GR without any non-standard additions that some have used to produce the variable light speed\cite{Mag3}.  We allow the metric that determines the curvature tensor to introduce the varying light speed.  This will create a relationship between the varying light speed and the components of the stress-energy tensor.  In order to use the standard GR action, we assume that $G/c^4 \equiv \hat{G}$ is constant.  This is needed to keep constant the Newtonian energy $-Gm_1 m_2 R$ when the rest energy of mass is $mc^2$.  We also assume that $\Lambda$ is  constant, possibly representing some kind of vacuum energy density.  We then 
 use the Lagrangian $L_{se}$ of the extended stress-energy tensor
\be  
S = \int \sqrt{-g}(R - 2\Lambda + 16\pi \hat{G} L_{se})d^4 \xi.
\ee 
 where $R$ is the Ricci scalar for the metric 
 \be
	ds^2 = g_{\mu \nu}d\xi^\mu d\xi^\nu,
 \ee
and $g$ is the determinate of $g_{\mu \nu}$.
  Minimizing the variation of $S$ with $g_{\mu \nu}$, we get the extended GR field equation:
\be
	G_{\mu \nu} + \Lambda g_{\mu \nu} = 8\pi \hat{G} T_{\mu \nu}.
\lbl{d5}
\ee

\section{Extended GR for homogeneous and isotropic universe}

We will now apply this field equation to an ideal fluid of density $\rho$ and pressure 
$p$ in a homogeneous and isotropic universe for which the extended FLRW line element 
in the variables $t,r,\theta,\phi$ is (eq \ref{1}):
\be
	ds^2 = c(t)^2dt^2 - a^2(\f {dr^2}{1-kr^2} + r^2d\theta^2 + r^2cos^2\theta d \phi^2).
\lbl{d5'}
\ee

As $ds$ is written in eq \ref{d5'}, the components of $G_{\mu \nu}$ will contain first and second derivatives of $c(t)$.
In order to find a solution to the field equation we will transform the time variable $t$ to $\xi^4 = \t$.  This will not change the relation of $G_{\mu \nu}$ to $T_{\mu \nu}$, but will eliminate the derivatives of $c(t)$ in $G_{\mu \nu}$ and transform $G_{\mu \nu}$ to a known solution.  %
For a perfect fluid of pressure $p$ and mass density $\rho$, we can define $\hat{\rho}\equiv \rho c^2$ so that $\hat{\rho}$ obeys the same conservation and acceleration laws using $d\t$ as does $\rho$ using $dt$ (Sect \ref{gr}).  Both $G_{\mu \nu}$ and $T_{\mu \nu}$ are \L covariant, so that the equivalency principle is maintained.  
We can then write the two significant field equations \cite[page 729]{MTW} for $a(\t)$ as
\be
	\f{3 \dot{a}^2}{a^2} + \f{3k}{a^2} - \Lambda = 8 \pi \hat{G} \hat{\rho} ,
\lbl{d6}
\ee
and
\be
	+2 \f{\ddot{a}}{a} +\f{\dot{a}^2}{a^2} +\f{k}{a^2} -\Lambda = -8 \pi \hat{G}p,
\lbl{d7}
\ee
where the dots represent derivatives with respect to $\hat{t}$.  All variables (including $\t$ and $a$) are in standard units. Eq \ref{d6} can be solved to give $a$ as a function of $\t. \rho, k$, and $\Lambda$.  When we know 
$c(\t)$, we can obtain the observables $a(t)$ and $c(t)$ by transforming $\t$ back to $t$. Solutions of these equations are carried out in \cite{RCF} for a particular $c(\t)$.  

Now, eq \ref{d6} can be multiplied by $a^3/3$, differentiated, and subtracted from $ \dot{a} a^2$ times eq \ref{d7} to give
\be
	\f{d}{d \t}(\hat{G} \rho a^3 c^2  ) = -\f{3G}{c^4}\dot{a} a^2 p  ,
\lbl{d8}
\ee
For small $p$, 
\be
	\hat{G} \rho a^3c^2 = \mbox{constant}
\lbl{d8'}
\ee
If the energy density consists of $n$ particles per unit volume of mass $m$, so $\rho = nm$, then the conservation of particles requires $na^3$ be constant (for small velocities).  This makes 
\be
	\f{Gm}{c^2} = \mbox{constant}.
\lbl{d8"}
\ee
This is consistent with our starting assumption.  

With constant $\hat{G}$, this has the same form as for $c=1$ and constant $G$.

\section{Summary}

We have here extended special and general relativity to include any variable light speed $c$ with a minimum of changes.  The extended Lorentz transform and extended Minkowski line element $ds$ are derived from the assumption that all moving observers at the same space-time point measure the same light speed, even though it may be variable.  The physical time $T$ often appears in the combination of $cdT$, so it convenient to use a differential construct $d\T$ in place of $dT$  where $d\T = cdT$.  We postulate that the rest mass $m$ of a particle varies so as to keep constant the rest mass energy $\hat{m} = mc^2$.  With the use of $d\T$ and $\hat{m}$ in particle interactions, we can generate SR velocity, energy-momentum, and force vectors whose components are covariant with the extended Lorentz transform.  A covariant stress-energy tensor $T^{\mu \nu} $ for an ideal fluid whose rest energy is conserved can also be generated with $d\T$ and constant $\hat{m}$.  A transform of electromagnetic (E/M) field variables assuming a constant fine structure constant and Rydberg frequency allows Maxwells equations and all the vectors and tensors of E/M theory to be preserved.  From the assumption of a universe that is homogeneous and isotropic in space we can generate an extended FLRW metric allowing for $c(t)$, where $t$ is the cosmic time appearing in the FLRW line element $ds$.  
We use the standard action for General Relativity, but use extended stress-energy tensors and allow $c,G$ to vary. However,  we assume $G/c^4 \equiv \hat{G}$ is constant in order to keep the Newtonian gravitational energy from varying with $c(t)$. This also keeps constant  
 the proportionality ratio in the field equation $\hat{G}$.
For systems in which $c$ is a function of the physical time $t$, we can use the a generalized time $\t = \int{c(t)dt}$ in the reduced curvature tensor $G^{\mu \nu}$  of the GR field theory and avoid having to introduce the first and second derivatives of $c(t)$.  In addition we can use all the relations between the curvilinear coordinates previously found for a constant $c=1$.  
We apply this to an ideal fluid in a homogeneous and isotropic universe, using sn extended FLRW metric, which yields equations for the scale factor $a(\t)$ of the universe in terms of $\t$.  For a particular $c(\t)$, these can be found as a function of observable time $t$ by the transform back from $\t$.

\section*{Acknowledgments} \lbl{ack}
\addcontentsline{toc}{section}{Acknowledgments}

I wish to acknowledge the invaluable help given by Paul Fife, University of Utah Mathematics Dept.  I also wish to thank David W. Bennett, Univ. of Utah Philosophy Dept, for his support and
encouragement and the faculty and facilities of the Physics and
Astronomy Department of Tufts University for the number of years
that I was allowed to visit there.  I thank Richard Price, Univ. of Utah Physics Dept,
for valuable discussions during the earlier part of this investigation, and Ramez Atiya for insightful discussions.  However, I alone am responsible for any errors of mathematics or interpretation that may be here.

\end{document}